# Acoustic one-way mode conversion and transmission by sonic crystal waveguides

Shiliang Ouyang[1], Hailong He[1], Zhaojian He[a)], Ke Deng[a)], and Heping Zhao

Department of Physics, Jishou University, Jishou 416000, Hunan, China

**Abstract:** We proposed a scheme to achieve one-way acoustic propagation and even–odd mode switching in two mutually perpendicular sonic crystal waveguides connected by a resonant cavity. The even mode in the entrance waveguide is able to switch to odd mode in the exit waveguide through a symmetry match between the cavity resonant modes and the waveguide modes. Conversely, the odd mode in the exit waveguide is unable to be converted into the even mode in the entrance waveguide as incident waves and eigenmodes are mismatched in their symmetries at the waveguide exit. This one-way mechanism can be applied to design an acoustic diode for acoustic integration devices and can be used as a convertor of the acoustic waveguide modes.

[1] Shiliang Ouyang and Hailong He contributed equally to this work.
[a)] Authors to whom correspondence should be addressed: hezj@jsu.edu.cn and dengke@jsu.edu.cn.

## I. INTRODUCTION

The recent progress of novel acoustic functionalities based on sonic crystals (SCs) has highlighted the possibility to realize compact on-chip solutions for various acoustic applications [1]. Such an integrated platform would consist of several elements of SC devices to achieve special sub-functions, e.g., beaming, focusing, and imaging of acoustic signals[2–5]. Alternatively, an efficient signal routing network, along which acoustic signals can be transmitted between different sub-functionalities, is also essential to really implement these on-chip solutions. Along this direction, acoustic waveguides based on defect/cavity modes or self-collimation effects in SCs have been thoroughly investigated[6–10].

In some areas such as nondestructive testing or medical imaging, developing one-way routing networks is very crucial to isolate different elements that allow acoustic signals to be manipulated independently of each other in the integrated platform [11–13]. Under these conditions, acoustic signals are transmitted in one direction only, which is somewhat similar to the energy-rectification characteristics of electric diodes. Indeed, recent years have witnessed an increasing research interest in unidirectional transmission of acoustic or elastic waves in consequence of high demands in applications[14–48]. For example, unidirectional transmission can be achieved by breaking time-reversal symmetry using nonlinear media[14–16, 34], resorting to magneto-acoustic effects[49], or using macroscopic flow fields[33, 38]. In addition, the breaking of spatial-inversion symmetry along opposite directions of incidence has also been widely discussed regarding unidirectional sound propagation, and essentially mode conversion between multiple scattering channels leads to asymmetric sound transmission in these systems [17–20]. Most assuredly, these achievements will provide enriched degree of freedom for single transmission in integrated acoustics.

In this paper, we propose a scheme to achieve one-way acoustic propagation and even–odd mode switching in two mutually perpendicular sonic crystal waveguides connected by a resonant cavity. The approach is based on symmetry matching and mismatching between the waveguide mode and cavity mode. The physical requirement is the breaking of space-reversal symmetry of the system. This scheme is believed to shed some light on the designing of one-way single transmission and even–odd mode switching for integrated acoustic applications. Compared with most existing methods to achieve unidirectional sound propagation, the proposed scheme can convert even mode inputs into odd mode outputs, thereby serving as an effective odd mode source, which is very useful in integrated acoustic applications. Here we note that[17, 36] even–odd conversion was also used to achieve one-way transmission of acoustic waves. Even–odd conversions were realized by introducing material modulations into the waveguide properties, which may lead to extra complexity in fabrication. Our scheme, which is based on SCs with line and point defects, offers simplified fabrication allowing devices to be more easily integrated with other SC based applications.

**II. METHODS**

Specifically, the acoustic system under study is a two-dimensional (2D) SC [Fig. 1(a)] formed as a square lattice of water cylinders (marked in light blue) immersed in mercury. Two mutually perpendicular waveguides are created by removing one row of cylinders in the SC. A resonant cavity, created to connect the two waveguides, is formed by an elliptical water cylinder (marked in red) surrounded by four water cylinders (marked in green). With the lattice period of the SC denoted by $a$, the cylinder radius is $0.3a$. The major axis and minor axis of the elliptical cylinder in the cavity are $0.6a$ and $0.2a$, respectively. The four surrounding cylinders in the cavity are displaced $0.2a$ from the center. Throughout the study, we use the finite-element method (FEM) of the commercial software platform COMSOL Multiphysics to perform band structure, transmission, and field distribution calculations. To perform the simulations, the lattice period is set as $a$=1mm, and the material parameters involved are as follows: $\rho$=13500 kg/m$^3$, $C_l$=1450 m/s for mercury; and $\rho$=1000 kg/m$^3$, $C_l$=1490 m/s for water, where $\rho$ and $C_l$ are the density and velocity, respectively, of the longitudinal component of sound for the associated medium. Water–mercury-based SCs have been widely adopted[50–58]. In practice, as suggested by many groups[51, 54, 56, 57], the water within the cylinders would be contained by means of some latex material. The mass density and speed of sound in rubber are comparable to those of water. Hence, for a sufficiently thin latex partition, the presence of this extra layer should not affect the calculations in any significant way and its effects can be neglected.

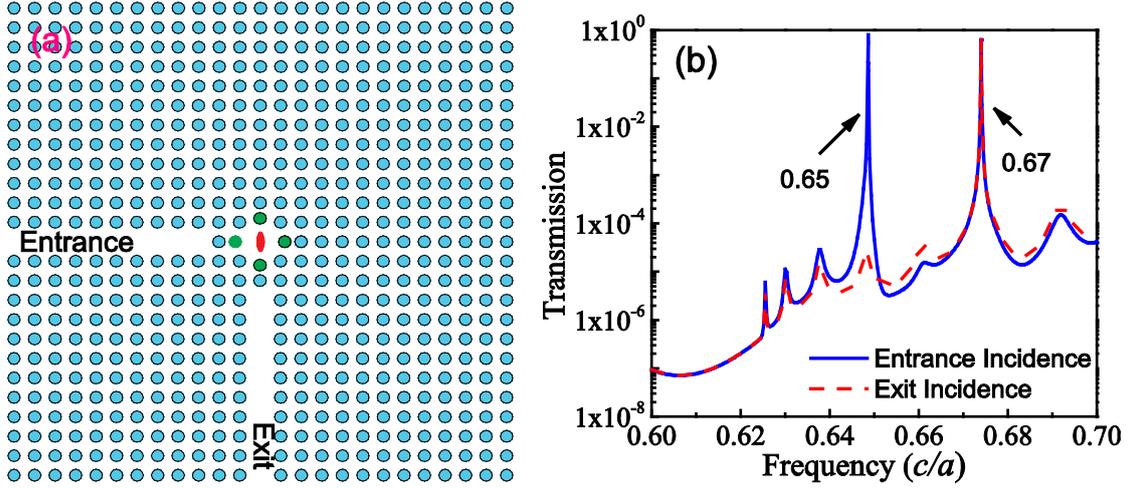

FIG. 1. (a) Schematic of the 2D sonic crystal with two mutually perpendicular waveguides (labeled Entrance and Exit) connected via a resonant cavity. (b) Transmission spectra of the acoustic waves through the system with incidence from the entrance waveguide (blue line) and exit waveguide (red dash line).

## III. RESULTS AND DISCUSSION

To demonstrate the one-way acoustic transmission through the composite system, we calculated the acoustic transmission of the system with a Gaussian beam incident from both the entrance and the exit waveguides [Fig. 1(a)]. From the results [Fig. 1(b)], we see two resonant transmission peaks at frequencies $0.65(c/a)$ and $0.67(c/a)$ for acoustic waves incident from the entrance waveguide; however, there is only one transmission peak at frequency $0.67(c/a)$ incident from the exit waveguide. Here $c$=1450 m/s is the acoustic velocity of background mercury. We find one-way acoustic transmission at frequency $0.65(c/a)$ with a large transmission contrast between entrance and exit incidences.

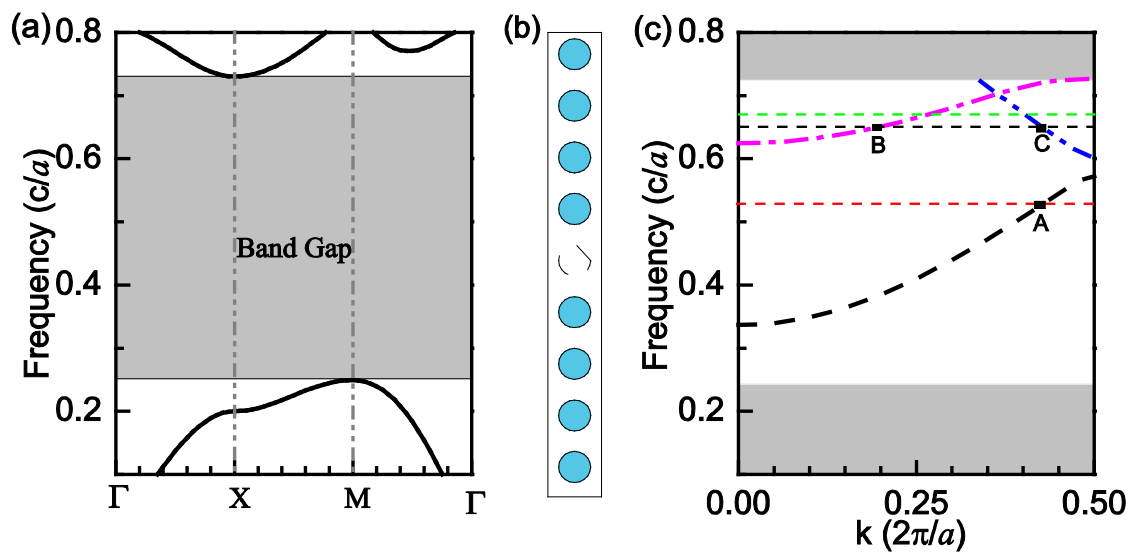

FIG. 2. (a) Band structure of a simple sonic crystal. (b) Supercell for band structure calculation of the waveguide. (c) Band structure of the waveguide.

To reveal the mechanism for the above one-way acoustic transmission through the combined system, we analyzed the dispersion properties of the SC-based waveguide structure and cavity structure, respectively. The band structure of a perfect SC was also calculated for reference [Fig. 2(a)], from which we see that a wide band gap is obtained between frequencies $0.25(c/a)$ and $0.73(c/a)$, benefitting the emergence of both waveguide modes and cavity modes. The waveguide structure is created by removing part of a row and a column of cylinders to form a line-defect in the SC, and its band structure is calculated by the supercell technique. The adopted supercell [Fig. 2(b)] used in our waveguide mode calculation has length $9a$ to avoid couplings between two neighboring defects; the band structure, [Fig. 2(c)] includes three waveguide bands within the gap of the perfect SC. Interestingly, between frequencies $0.62(c/a)$ and $0.73(c/a)$, there are two crossed waveguide bands. To further explore the properties of these waveguide modes, we investigated their eigenfield patterns. We exemplify the field characteristics using three typical waveguide modes [labeled A, B, and C in Fig. 2(c)]. The field patterns [Fig. 3(a)–(c)] show that A and C are even modes as their field pattern are symmetrical with respect to the waveguide direction (the X-axis direction in Fig. 3), and B is an odd mode as its field pattern is anti-symmetrical with respect to the waveguide direction. Hence, for the frequencies between $0.62(c/a)$ and $0.73(c/a)$, the even modes and odd modes coexist in the waveguide but have different wave numbers. This is the key point governing the one-way acoustic transmission in our proposed system that is analyzed in detail below.

Next we investigate the dispersion properties of the cavity structure. The cavity with its central elliptical cylinder surrounded by four cylinders has the adopted supercell in our cavity mode calculation. Fig. 4(a)]; with verified cavity modes [horizontal dash lines in Fig. 2(b)]. There exist three cavity modes within the band gap of the perfect SC, located at frequency 0.54($c/a$), the frequency of mode A, 0.65(c/a) and 0.67(c/a), the frequencies of modes B and C respectively. The eigen-field patterns of these three cavity modes were calculated [Fig. 4(b)–(d)], respectively. Since the center cylinder is elliptical, there are two symmetry axes in the field patterns as expected. One can see that, for the cavity mode at frequency 0.54($c/a$), the field pattern is symmetrical with respect to the major axis of the ellipse (the Y-axis direction) but anti-symmetric with respect to the minor axis (X-axis direction); see Fig. 4(b); While for the mode at frequency 0.65($c/a$), the situation is reversed as the field pattern is symmetrical to minor axis but anti-symmetrical to major axis [Fig. 4(c)]; For the higher frequency mode at 0.67($c/a$), the field pattern is symmetric for both the major and minor axis [Fig. 4(d)]. As mentioned above, the two intercrossed waveguide bands [Fig. 2(b)] play a key role in governing the one-way mode acoustic transmission. Here, the cavity mode with frequency 0.53($c/a$) falls out of the two intercrossed bands [Fig. 2(c)], and henceforth will not be considered.

We return to the one-way transmission of the compound system [Fig. 1(a)]. According to our analysis in the previous paragraph, for the designed compound system, the field pattern of cavity mode at frequency 0.65($c/a$) is symmetric with respect to the entrance waveguide but anti-symmetrical for the exit waveguide. Therefore, for acoustic waves incident from the entrance, the excited even (symmetric) mode in the entrance waveguide can be successfully converted into the odd (anti-symmetric) mode in the exit waveguide by exciting the cavity mode because in this instance the symmetry matches. However, for acoustic waves incident from the exit, the excited even mode (symmetric) in the exit waveguide will be blocked by the resonant cavity, because the waveguide mode no longer matches the symmetry of the cavity mode. This give rise to the one-way transmission observed in Fig. 1(b). In addition, it can be inferred that an effective one-way mode-conversion from even type to odd type can also be realized by our proposed structure. Such an even–odd mode conversion has been extensively investigated in integrated optics, because it plays a very important role in increasing transmission capacity by carrying information on different modes with only a single wavelength carrier[59–64]. This mode conversion concept can be used in many applications such as subwavelength edge detection[65], cavity biosensors[66], and direction selective structures[67]. Moreover, this mode conversion scheme provides an effective way to generate acoustic source with odd symmetry, which may find many applications in integrated acoustics[68, 69].

To verify these conclusions, we simulated the field distribution for this one-way transmission and mode-conversion effect. From the results [Fig. 5(a) and (b)], for an entrance incidence (indicated by horizontal blue arrow), the excited even mode in the entrance waveguide is successfully converted into the odd mode in the exit waveguide by exciting the cavity mode as expected. For an exit incidence (indicated by vertical red arrow), one sees that the excited even mode in the exit waveguide is blocked by the cavity because of the symmetry mismatch discussed above. For comparison, the field distributions of the system with acoustic waves of frequency $0.67(c/a)$ incident from entrance and exit waveguides were also simulated [Fig. 5(c) and (d), respectively]. We find that, as the cavity mode is symmetric (even) for both entrance and exit waveguides [Fig. 4(d)], the excited even mode in the entrance (exit) waveguide is converted into an even mode in the exit (entrance) waveguide through the excitation of the cavity mode. Therefore, acoustic waves at frequency $0.67(c/a)$ can be transmitted through the system for both entrance incidence and exit incidence, in agreement with the transmission results of Fig. 1(b).

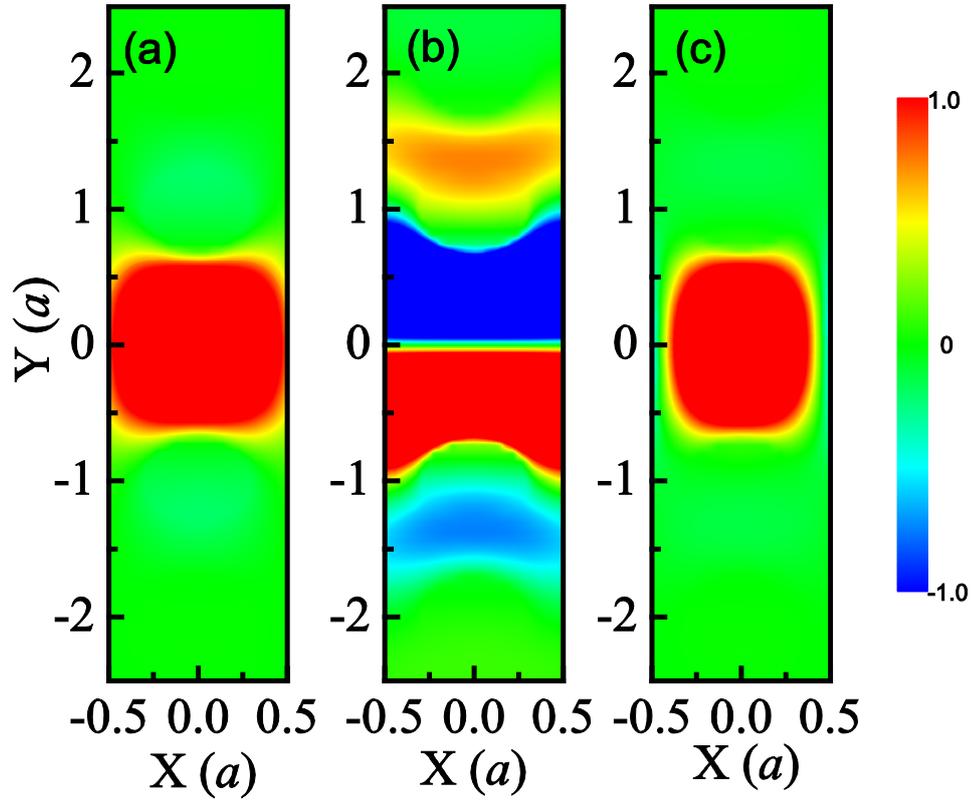

FIG. 3. Eigen-pressure fields of the waveguide modes for (a) f=0.53(*c*/a), k=0.434(2π/*a*); (b) f=0.65(*c*/a), k=0.22(2π/*a*); and (c) f=0.65(*c*/a), k=0.426(2π/*a*), corresponding to modes labeled A, B, and C respectively, in Fig. 2(c). Red/blue regions represent positive/negative pressure fields.

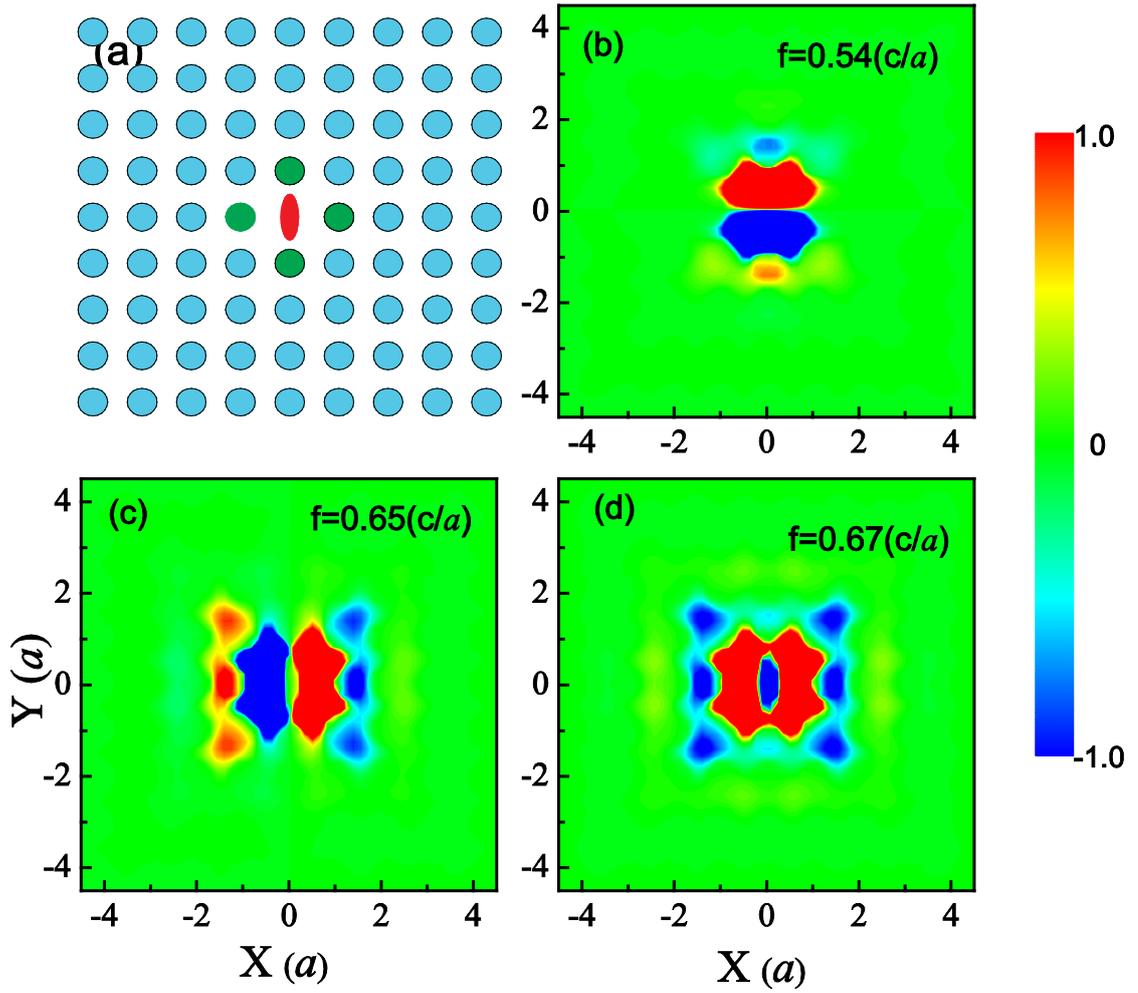

FIG. 4. (a) Schematic of the 9×9 supercell for the resonant cavity; (b)–(d) eigen-pressure fields of the resonant cavity modes with frequency f=0.54(*c/a*), f=0.65(*c/a*), and f=0.67(*c/a*), respectively. Red/blue regions represent positive/negative pressures.

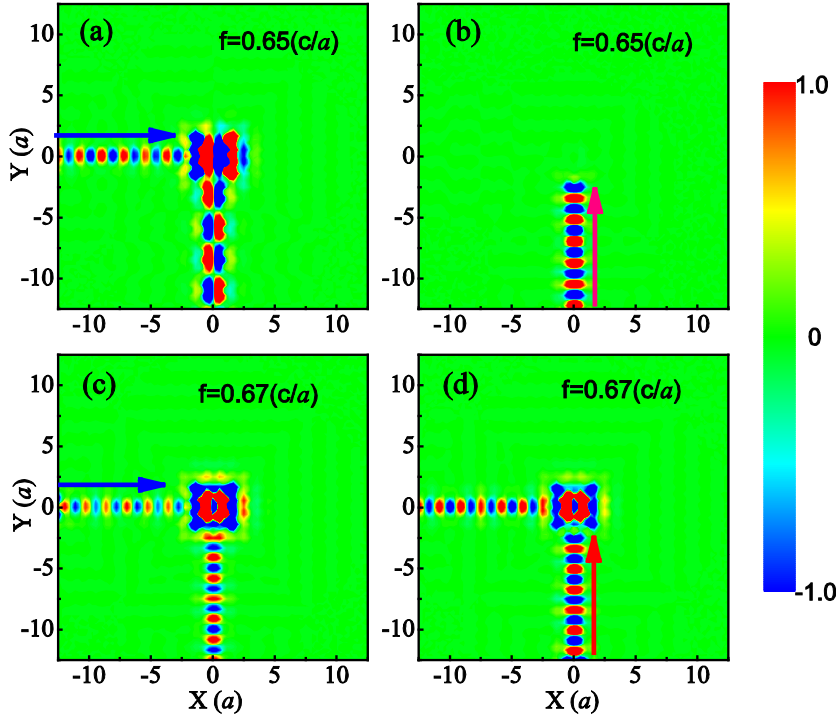

FIG. 5. Pressure field distributions of the composite system under cavity resonances with two higher frequencies with different acoustic incidences: (a), (c) for frequency f=0.65(*c/a*) with entrance incidence and exit incidence. (b), (d) for frequency f=0.67(*c/a*) with entrance incidence and exit incidence.

## IV. CONCLUSION

We have studied the one-way waveguide-mode conversion and acoustic transmission in two mutually perpendicular sonic crystal waveguides connected by an asymmetric resonant cavity. The physical feature enabling this phenomenon is the breaking of space-reversal symmetry of the system. Specifically, the acoustic one-way properties stem from the symmetry matching of the waveguide mode and resonant cavity mode. This one-way mechanism can be applied in designing an acoustic diode for acoustic integration devices and can be used to convert acoustic modes in the waveguide, providing a way to design anti-symmetric acoustic sources. In addition, because the device is sensitive to the acoustic frequency, it can also be used for frequency filtering. The mechanism can be generalized to EM and elastic waves.

**ACKNOWLEDGEMENTS**

This work is supported by the National Natural Science Foundation of China (Grant Nos. 11464012, 11304119, 11564012, and 11564013), Natural Science Foundation of Hunan province, China (Grant No. 2016JJ2100), and the Aid Program for Science and Technology Innovative Research Team in Higher Educational Institutions of Hunan Province.